\documentclass[conference]{IEEEtran}
\usepackage[utf8]{inputenc}

\usepackage{amsmath}
\usepackage{amsfonts}
\usepackage{amssymb}

\usepackage{graphicx}
\graphicspath{ {./graphics/} }
\usepackage{url}

\usepackage{xcolor}

\usepackage{booktabs}
\usepackage{arydshln}

\usepackage[ruled,lined,linesnumbered]{algorithm2e}

\def \calG {\mathcal{G}}
\def \calN {\mathcal{N}}
\def \calE {\mathcal{E}}

\title{A Study of Age and Gender seen through Mobile Phone Usage Patterns in Mexico}

\author{

\IEEEauthorblockN{Carlos Sarraute}
\IEEEauthorblockA{Grandata Labs, Argentina \\
\texttt{ \small {charles@grandata.com} } }

\and

\IEEEauthorblockN{Pablo Blanc}
\IEEEauthorblockA{Mathematics Dept., FCEN, UBA \\
\texttt{ \small {pblanc@dm.uba.ar} } }

\and

\IEEEauthorblockN{Javier Burroni}
\IEEEauthorblockA{Grandata Labs, Argentina \\
\texttt{ \small {javier.burroni@grandata.com} } }

}

\begin{document}
\maketitle

\begin{abstract}

Mobile phone usage provides a wealth of information,
which can be used to better understand the demographic
structure of a population. In this paper we focus on the
population of Mexican mobile phone users.
Our first contribution is an observational study of mobile phone
usage according to gender and age groups.
We were able to detect significant differences in phone usage among 
different subgroups of the population.
Our second contribution is to provide a novel methodology
to predict demographic features (namely age and gender) of unlabeled
users by leveraging individual calling patterns, as well as the
structure of the communication graph.
We provide details of the methodology and show experimental results
on a real world dataset that involves millions of users.

\end{abstract}

\section{Introduction}

Mobile phones have become prevalent in all parts of the world,
in developed as well as developing countries, and
provide an unprecedented source of information on the 
dynamics of the population on a national scale.
In particular, mobile phone usage is starting to be used to
perform quantitative analysis of the demographics of the population,
respect to key variables such as gender, age, level of education
and socioeconomic status (for example see \cite{blumenstock2010mobile,blumenstock2010s}).

In this work we combine two sources of information: 
transaction logs from a major mobile operator in Mexico,
and information on the age and gender of a subset of
the population.
This allows us to perform an observational study of 
mobile phone usage, differentiated by gender and age groups.
This study is interesting in its own right, 
since it provides knowledge on the structure and
demographics of the mobile phone market in Mexico.
We can start to fill gaps in our understanding of basic demographic questions:
Are inequalities between men and women, as reported by \cite{katz_economics_2001},
reflected in mobile phone usage (in calling and texting patterns)?
What are the differences in mobile phone usage between different age ranges?
 
The second contribution of this work is to apply 
the knowledge on calling patterns to predict demographic features, 
namely to predict the age and gender of unlabeled users.
We present methods that rely on individual calling patterns,
and introduce a novel algorithm that exploits the structure of the social graph (induced by communications),
in order to improve the accuracy of our predictions.

Being able to understand and predict demographic features such as age and
gender has numerous applications, 
from market research and segmentation to the 
possibility of targeted campaigns (such as health campaigns for women \cite{frias2010gender}).

The remainder of the paper is organized as follows:
section~\ref{sec:dataset} provides an overview of the datasets that we used in 
this study. 
Section~\ref{sec:observations} describes the observations that we gathered,
the insights gained from data analysis, and
the differences that could be seen in CDR features between genders and age groups.
In particular, very clear correlations have been observed in the links between users
according to their age.
In section~\ref{sec:identification} we present the models that we used to identify
the age and gender of unlabeled users.
We show the experimental results obtained using classical Machine Learning techniques
based on individual attributes, both for gender and age.
We introduce a novel algorithm that leverages the links between users
both in its pure graph based form (section~\ref{sec:age-identification-links}),
and combined form (section~\ref{sec:age-identification-combined}).
The results of our experiments show that the pure graph based algorithm has the best
predictive power.
Section~\ref{sec:conclusion} concludes the paper with ideas for future work.

\section{Dataset Description} \label{sec:dataset}

The dataset used for this study consists of cell phone call 
and SMS (Short Message Service) records
collected in Mexico for a period of $M$ months ($M = 3$) by a large mobile phone operator.
The dataset is anonymized.
For our purposes, each CDR (Call Detail Record) is represented as a tuple 
$\left < x, y, t, dur, d, l \right >$,
where $x$ and $y$ are the encrypted phone numbers of the caller and the callee,
$t$ is the date and time of the call,
$dur$ is the duration of the call,
$d$ is the direction of the call (incoming or outgoing, with respect to the mobile operator client),
and $l$ is the location of the tower that routed the communication.
Similarly, each SMS record is represented as a tuple $\left < x, y, t, d \right >$.

We construct a social graph $\calG = < \calN_T, \calE > $,
based on the aggregated traffic of $M$ months.
We use $\calN_T$ to denote the set of mobile phone users that appear in the dataset.
$\calN_T$ contains about 90 million unique cell phone numbers.
Among the numbers that appear in $\calN_T$, only some of them are clients of the 
mobile phone operator: we denote that set $\calN_O$.

For this study, we had access to basic demographic information
for a subset of the operator clients, that we denote $\calN_{GT}$ 
(where \emph{GT} stands for \emph{ground truth}).
The size of this labeled set $| \calN_{GT} | $ is about 500,000 users.
The following relation holds between the three sets: $\calN_{GT} \subset \calN_O \subset \calN_T$.

\begin{figure}[ht]
	\centering
	{\includegraphics[trim=1.2cm 0.3cm 1.2cm 1.2cm, clip=true, width=0.98\linewidth]{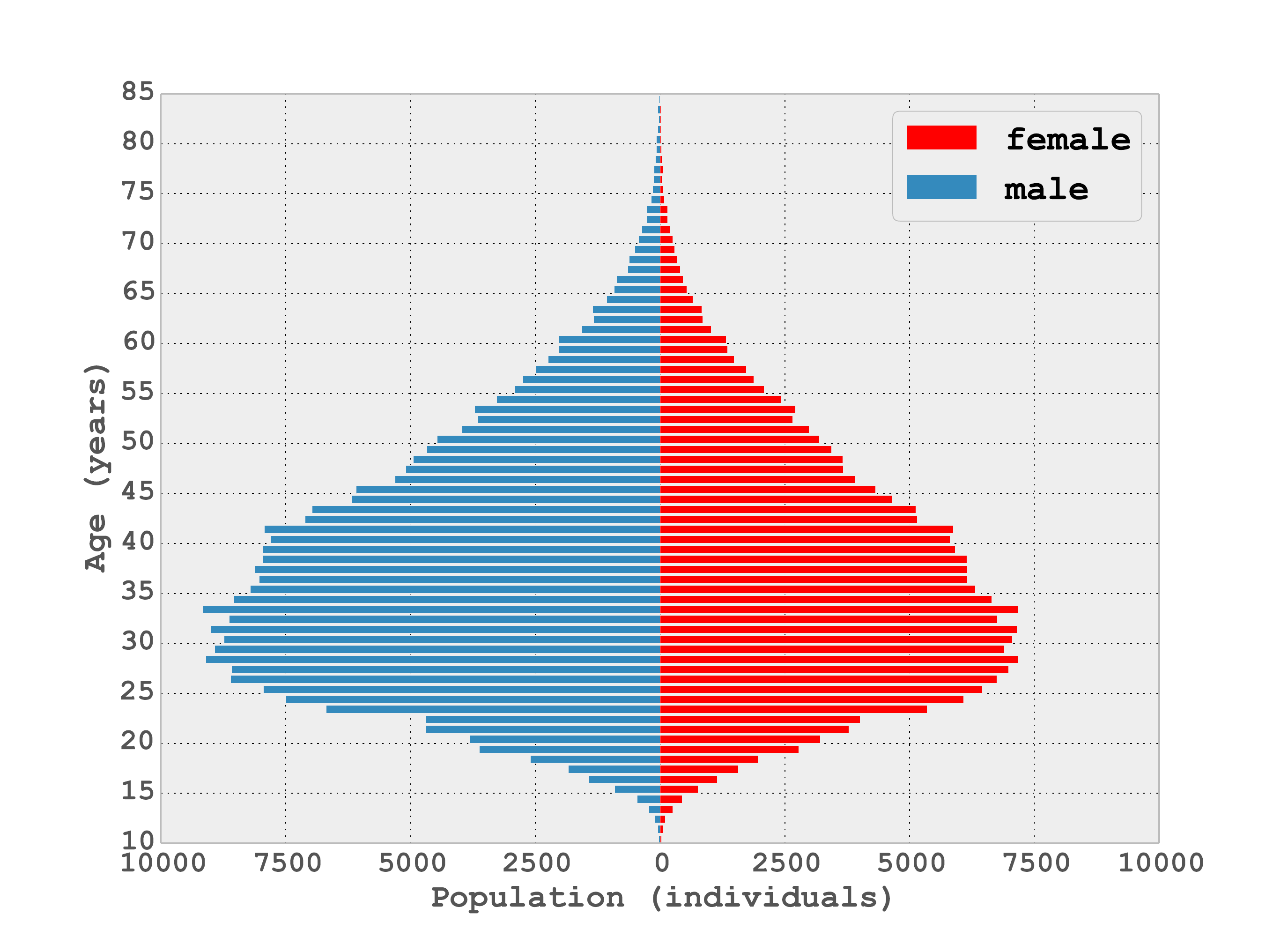}}
	\caption{Pyramid of ages of the labeled set $\calN_{GT}$, showing the number of
	individuals for each age value (in years).}
	\label{fig:pyramid}
\end{figure}

Fig.~\ref{fig:pyramid} shows the pyramid of ages of the labeled set.
This pyramid is different from the age pyramid of the entire population,
since it only contains mobile phone users (that pertain to $\calN_{GT}$).
Some basic observations on this pyramid: there are more men ($56.83\%$)
than women ($43.17\%$) in the labeled set.
The mean age is $37.23$ years for men and $36.47$ years for women.

\section{Observational Study} \label{sec:observations}

We report in this section the quantitative analyses that we performed on the dataset,
which provide the basis for the prediction algorithms 
(which will be described in section \ref{sec:identification}).
Our raw data input are the transaction logs (that contain billions of records at the 
scale of Mexico). A first step in the study was 
to generate characterization variables for each user, which
summarize their individual and social behavior.
We also describe the preprocessing performed on the data,
the key features identified with 
PCA (Principal Component Analysis),
and the differences observed. 
We will focus on some examples to illustrate the kind of observations we obtained for many variables.
Statistically significant differences have been found, which
motivates our attempt to identify gender and age based on communication patterns.

\subsection{Characterization Variables} \label{sec:variables}

In this study we chose to characterize the users in $\calN_O$ (i.e. clients of the
mobile operator), since for users in $\calN_T \setminus \calN_O$ we can 
only see part of their calls and messages (those exchanged with users in $\calN_O$),
thus including them would require a different calibration.
For users in $\calN_O$, we computed the following variables which characterize their
calling consumption behavior (also called ``behavioral variables" in \cite{frias2010gender}).

\begin{itemize}
\item \textit{Number of Calls}. We consider incoming calls i.e.,
the total number of calls received by user $u$ during a period
of three months, as well as outgoing calls i.e., total number of
calls made by user $u$.
Additionally, we distinguish whether those calls happened
during the weekdays (Monday to Friday) or during the weekend; 
and we further split the weekdays in two parts: the ``daylight'' (from 7 a.m. to 7 p.m.)
and the ``night'' (before 7 a.m. and after 7 p.m.).

We thus have $3 \times 4$ variables for the number of calls, given by the
Cartesian product: $ [ in, out, all ] \times [weekdaylight, weeknight, weekend, total] $.

\item \textit{Duration of Calls}. We calculate the total duration 
of incoming calls and outgoing calls of user $u$ during the period of three months. 
As before, we distinguish between weekdays (by daylight and by night) and weekends,
to get a total of 12 variables for the duration of calls.

\item \textit{Number of SMS}. We consider incoming messages 
(received by user $u$)
and outgoing messages (sent by user $u$).
Similarly we distinguish between weekdays (by daylight and by night) and weekends,
to get a total of 12 variables for the number of SMS.

\item \textit{Number of Contact Days}. We consider the number of days where the user has activity. We distinguish between calls and SMS, and between incoming, outgoing or any activity. This way we get 6 variables related to the number of activity days. 

\end{itemize}

We also computed variables which characterize the social network of users
based on their use of the cell phone (also called ``social variables'' in \cite{frias2010gender}).

\begin{itemize}
\item \textit{In/Out-degree of the Social Network}: The in-degree for
user $u$ is the number of different phone numbers
that called or sent an SMS to that user. 
The out-degree is the number of distinct phone numbers
contacted by user $u$.

\item \textit{Degree of the Social Network}: 
The degree is the number of unique phone numbers
that have either contacted or been contacted by user
$u$ (via voice or SMS).

\end{itemize}

\subsection{Data Preprocessing} \label{sec:preprocessing}

Many of the variables that we generated have a right skewed or heavy tailed distribution.
Our experiments showed that this skewness affects the results
given by Machine Learning algorithms (described in section \ref{sec:gender-identification}).
Therefore as part
of the data preprocessing we also considered the logarithmic version of the variables.
We discuss this preprocessing in more detail for one variable, that we 
use as running example: \emph{in-time-total}, i.e. the total duration of incoming calls
for a given user.

As can be seen in Table~\ref{tab:prelog} (left), the quartiles of the variable \emph{in-time-total} 
lie in different orders of magnitude, 
in particular the ratio ${IQR}/{Q_2} = {(Q_3-Q_1)}/{Q_2}$ is well above $1$.

\begin{figure}[ht]
	\centering
	\includegraphics[clip, trim=0.2cm 1cm 0cm 0cm, width=0.85\linewidth]{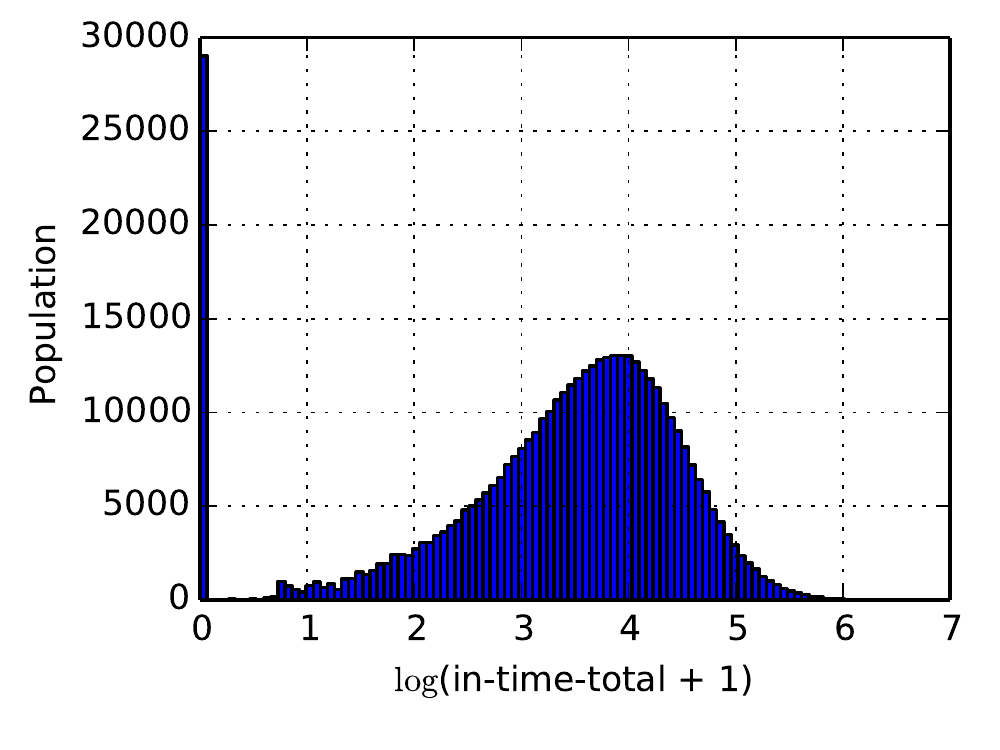}
	\caption{Histogram of $\log(\textit{in-time-total} + 1)$. 
The distribution can easily be seen as a mixture between a distribution with all its density in $0$ 
(users that have no incoming calls)
and a bell shaped distribution for $\log(\textit{in-time-total} + 1) > 0$.
}
	\label{fig:logintimetotal}
\end{figure}

To improve the results given by the Machine Learning methods,
we transform the data using the function
$ T(x) = \log_{10}(x + 1) $.
After the transformation, we found the statistics in Table~\ref{tab:prelog} (right). 
As we can see, the quartiles are in the same order of magnitude, 
and the ratio ${IQR}/{Q_2} $ is below $1$. 
The resulting distribution is shown in Fig.~\ref{fig:logintimetotal}.

\begin{table}[ht]
\caption{Statistic summary for \textit{in-time-total} and its logarithmic transformation}
\label{tab:prelog}

\centering
\begin{small}
\begin{tabular}{l r r}
\toprule
{}    &  \textit{in-time-total} (seconds) & $\log(\textit{in-time-total} + 1)$ \\
\midrule
count &   131770.00 &  131770.00 \\
mean  &    16239.28 &       3.31 \\
std   &    50023.16 &       1.23 \\
min   &        0.00 &       0.00 \\
25\%  &      662.00 &       2.82 \\
50\%  &     3838.00 &       3.58 \\
75\%  &    14108.00 &       4.14 \\
max   &  4045686.00 &       6.60 \\
\bottomrule
\end{tabular}
\end{small}

\end{table}

In conclusion, we decided to include both plain variables as well as their logarithmic values,
and let our Machine Learning algorithms select which variables are most relevant
for modeling a given target variable (e.g. gender and age).
We also rescaled all variables to take values between 0 and 1.

\subsection{Insights on Key Features from PCA} \label{sec:pca}

We performed PCA (Principal Component Analysis) on the characterization variables,
in order to gain information on which are the most important variables.
This gave us interesting insights on the key features of the data.
We describe the first 4 eigenvectors (which account for $89.6\%$ of the variance).

The first eigenvector retains $76.0\%$ of the total variance.
This eigenvector is dominated by the logarithmic 
version\footnote{We note that in all eigenvectors, the logarithmic version of the variables got systematically higher coefficients than the plain variables.}
of the total number of calls, total duration of calls and total number of SMS.
This result shows that the level of activity of users exhibits the highest variability,
and therefore is a good candidate to characterize users' social behavior.

The second eigenvector, which retains $6.7\%$ of the variance,
gives high positive coefficients to ``outgoing'' variables
(number of outgoing calls, duration of outgoing calls, number of SMS sent) and negative coefficients to  
``incoming'' variables (number of incoming calls, duration of incoming calls, number of SMS received). 
This suggests that the difference of outgoing minus incoming 
communications is also a good variable to describe users' social behavior.

The third eigenvector (with $4.4\%$ of the variance) gives positive coefficients
to the ``voice call'' variables, and negative coefficients to the ``SMS'' variables
(intuitively the difference between voice and SMS usage is relevant).

The fourth eigenvector (with $2.5\%$ of the variance) gives positive coefficients
to the ``weeknight'' and ``weekend'' variables (communications made non-working hours, i.e. during the night or during the weekend), and negative coefficients to ``weeklight'' variables (communications made during the day, from Monday to Friday, which correspond roughly to working hours).

\subsection{Observed Gender Differences} \label{sec:observations-gender}

We report in Table~\ref{tab:mean-male-female} the average of 6 key variables, 
distinguished by gender. 
We know from PCA that the number of calls,
and the total duration of calls made by a user, characterize their level of activity.
A natural question is whether differences can be observed between genders.
We found that for all the variables, there is a significant difference between genders,
with a very small $p$-value ($p < 10^{-10}$).
Recall that these values are computed as the aggregation of calls during a period
of $M = 3$ months for users in $\calN_{GT}$ (about 500,000 users).

\begin{table}[ht]
\caption{Sample mean of key variables for female and male users. 
The durations are expressed in seconds.
}
\label{tab:mean-male-female}

\centering
\begin{small}
\begin{tabular}{ l r r }
\toprule
Variable & Female & Male \\
\midrule
$\hat{\mu}$(total duration)          & 10038.75 & 10663.17 \\
$\hat{\mu}$(total duration outgoing) &  6359.96 &  7239.53 \\
$\hat{\mu}$(total duration incoming) &  3678.78 &  3423.64 \\
\midrule
$\hat{\mu}$(number of calls total)    & 72.847  & 81.348 \\
$\hat{\mu}$(number of calls outgoing) & 44.136  & 50.047 \\
$\hat{\mu}$(number of calls incoming) & 28.710  & 31.301 \\
\bottomrule
\end{tabular}
\end{small}

\end{table}

Table~\ref{tab:mean-male-female} shows that men have on average
higher total number of calls, and total duration of calls (measured in seconds).
However an interesting pattern can be seen when we distinguish incoming
and outgoing calls: the duration of outgoing calls is higher for men,
but the duration of incoming calls is higher for women.
It follows that the net duration of calls (the difference between outgoing and incoming calls)
has a marked gender difference: 
the sample mean $\hat{\mu}$(net total duration) = 3815.88 seconds for men,
and $\hat{\mu}$(net total duration) = 2681.18 seconds for women.
We note that the number of outgoing calls is higher than the number of incoming calls
for both men and women, due to a particularity of our dataset
(for all the users in $\calN_T$ the total number of incoming and outgoing calls is
the same, but for users in $\calN_{GT}$ there is a higher proportion of outgoing calls).

We also compute the conditional probability $p(g'|g)$ that a random call made by an 
individual with gender $g$ has a recipient with gender $g'$, 
where we denote male by $M$ and female by $F$. 
For the calls originated by male users, we found that 
$p(F|M) = 0.3735$ and $p(M|M) = 0.6265$. 
For the calls originated by female users, we found that
$p(F|F) = 0.4732$ and $p(M|F) = 0.5268$.
We can see a difference between genders, 
in particular men tend to talk more with men, and women tend to talk more with women.
More precisely:
\begin{equation}
\label{eq:gender-homophily}
\begin{split}
p(M|F) < p(M) &= 0.5683 < p(M|M) \\
p(F|M) < p(F) &= 0.4317 < p(F|F)
\end{split}
\end{equation}
Similar observations have been made in the case of the Facebook social graph~\cite{ugander5anatomy}.

\subsection{Observed Age Differences} \label{sec:observations-age}

We approached the study of mobile phone usage patterns according to age
by dividing the population in $C=4$ categories: 
below 25 years, from 25 to 34 years, from 35 to 49 years, and 50 years or above. 
We use this same structure for age prediction 
(in section~\ref{sec:age-identification-node}).

Since we are dealing with more than $2$ groups, comparing differences between groups 
requires using the correct tool, as the probability of making a type I error (null hypothesis incorrectly rejected) increases. 
In order to compare the means (of the $\log$) of the variables for each age group, 
we conduct a {Tukey's HSD} (Honest Significant 
Difference) test. 
This method tests all groups, pairwise, simultaneously.  
We found a list of 20 variables for which the null hypothesis of same mean ($H_0$) is rejected for all pair of groups, i.e. $\mu_i \neq \mu_j$ for every $i \neq j$.

\begin{table}[ht]
\caption{Tukey HSD for the variable $\log_{10}(\textit{in-time-total} + 1)$
}
\label{tab:tukeyHSD}

\centering
\begin{small}
\begin{tabular}{cccccc}
\toprule
group1 & group2 & meandiff & lower & upper & reject  \\
\midrule
       0        &        1        &       0.1567      &     0.1328     &     0.1807     &       True       \\
       0        &        2        &       0.1326      &     0.1088     &     0.1564     &       True       \\
       0        &        3        &       0.2367      &     0.2122     &     0.2612     &       True       \\
       1        &        2        &      -0.0242      &    -0.0407     &    -0.0076     &       True       \\
       1        &        3        &        0.08       &     0.0625     &     0.0975     &       True       \\
       2        &        3        &       0.1041      &     0.0868     &     0.1214     &       True       \\
\bottomrule
\end{tabular}
\end{small}

\end{table}

We illustrate the difference between age groups for our running example: \emph{in-time-total} (total duration of incoming calls per user). 
Table~\ref{tab:tukeyHSD} show the result of Tukey HSD (where FWER=0.05) for the variable 
$\log_{10}(\textit{in-time-total} + 1)$, obtained after the preprocessing step.
The 4 age groups are labeled 0, 1, 2, 3. 
Pairwise comparisons are done for all combinations (of group1 and group2). The null hypothesis is rejected for all pairs;
in other words, all the groups are found statistically different respect to this variable.

\begin{figure}[th]
\centering
\begin{footnotesize}

	\begin{minipage}{.48\linewidth}
		\centering
		\includegraphics[width=\textwidth]{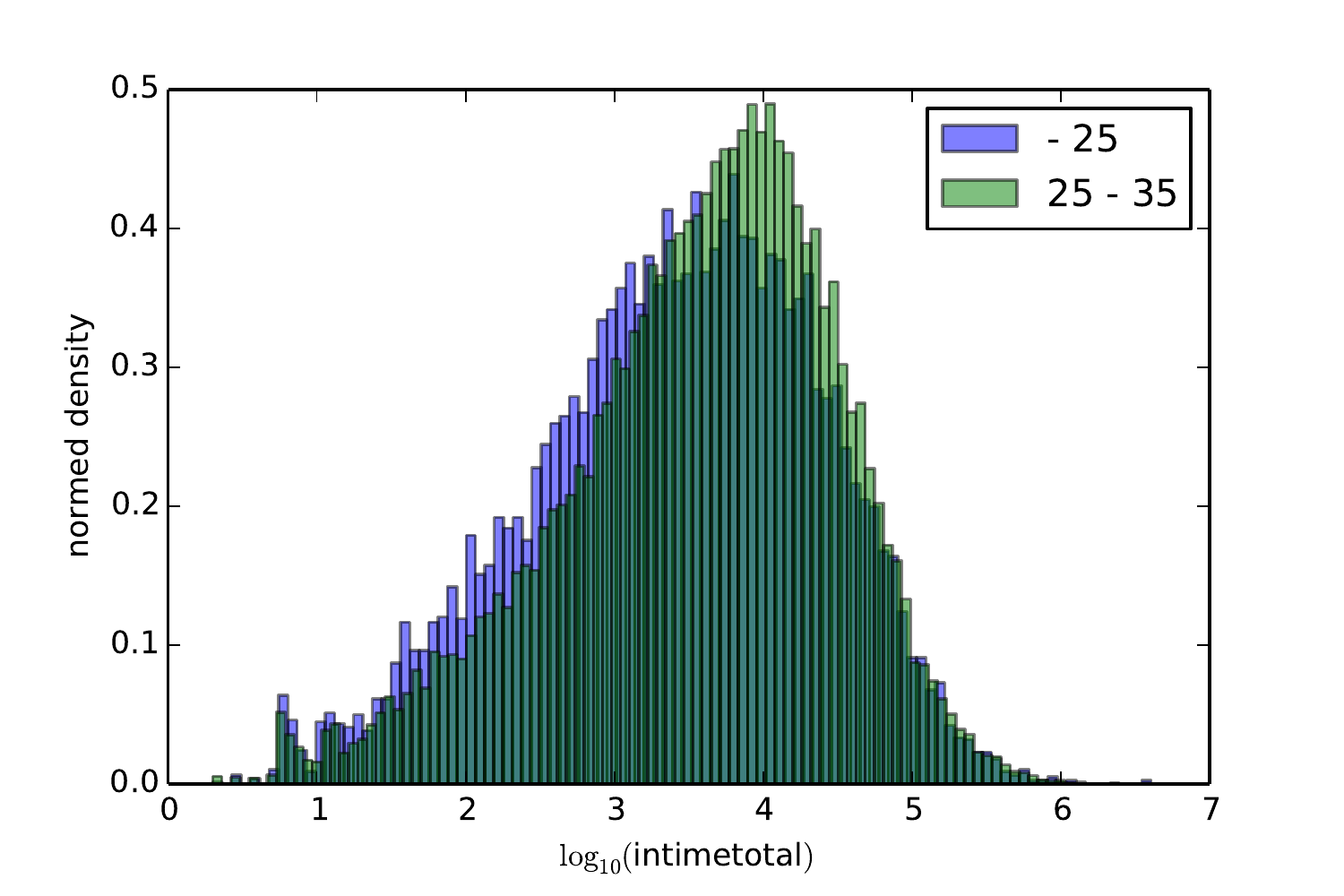}
		(a) 
	\end{minipage}
	\begin{minipage}{.48\linewidth}
		\centering
		\includegraphics[width=\textwidth]{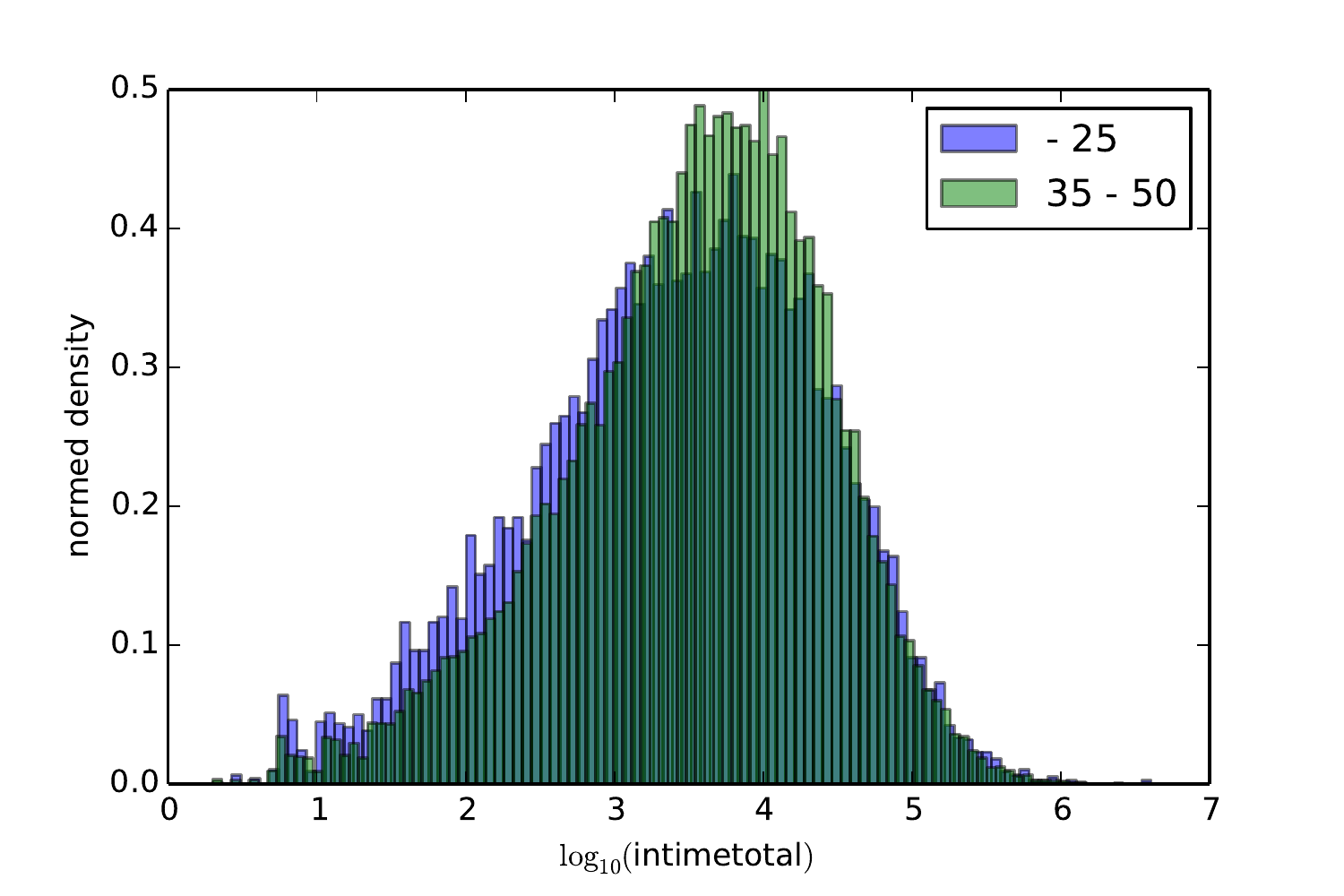}
		(b) 
	\end{minipage}
	\begin{minipage}{.48\linewidth}
		\centering
		\includegraphics[width=\textwidth]{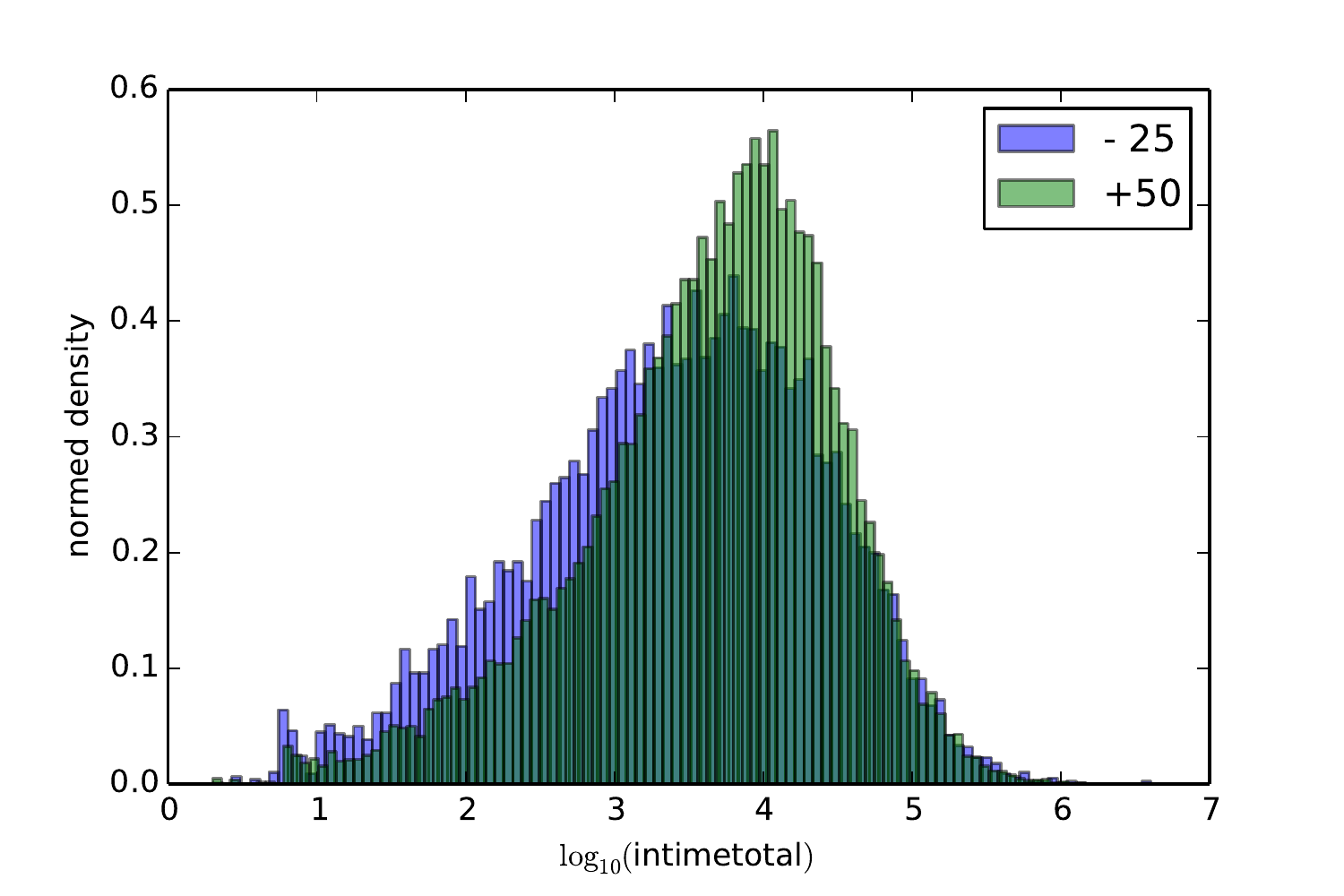}
		(c) 
	\end{minipage}
	\begin{minipage}{.48\linewidth}
		\centering
		\includegraphics[width=\textwidth]{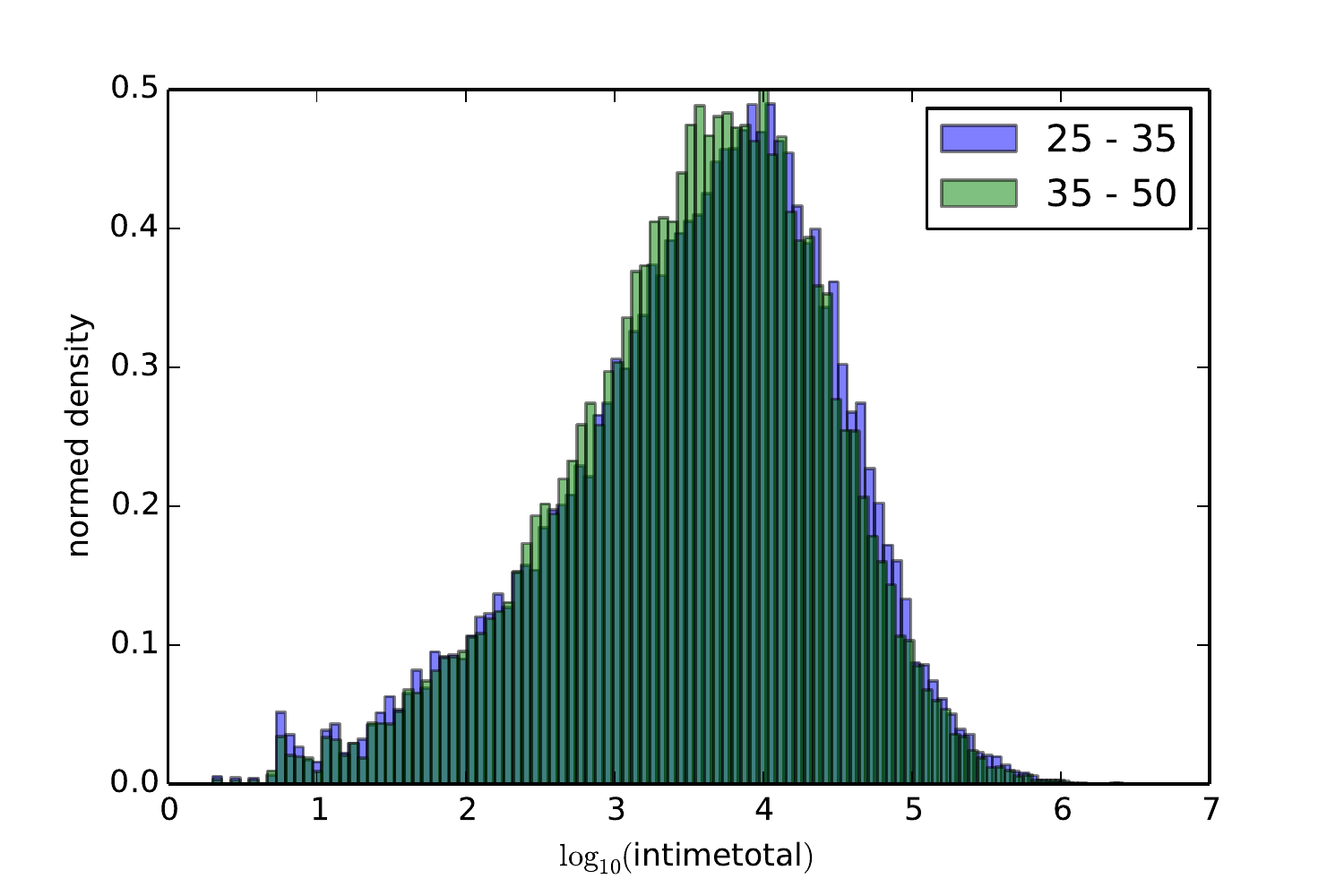}
		(d) 
	\end{minipage}
	\begin{minipage}{.48\linewidth}
		\centering
		\includegraphics[width=\textwidth]{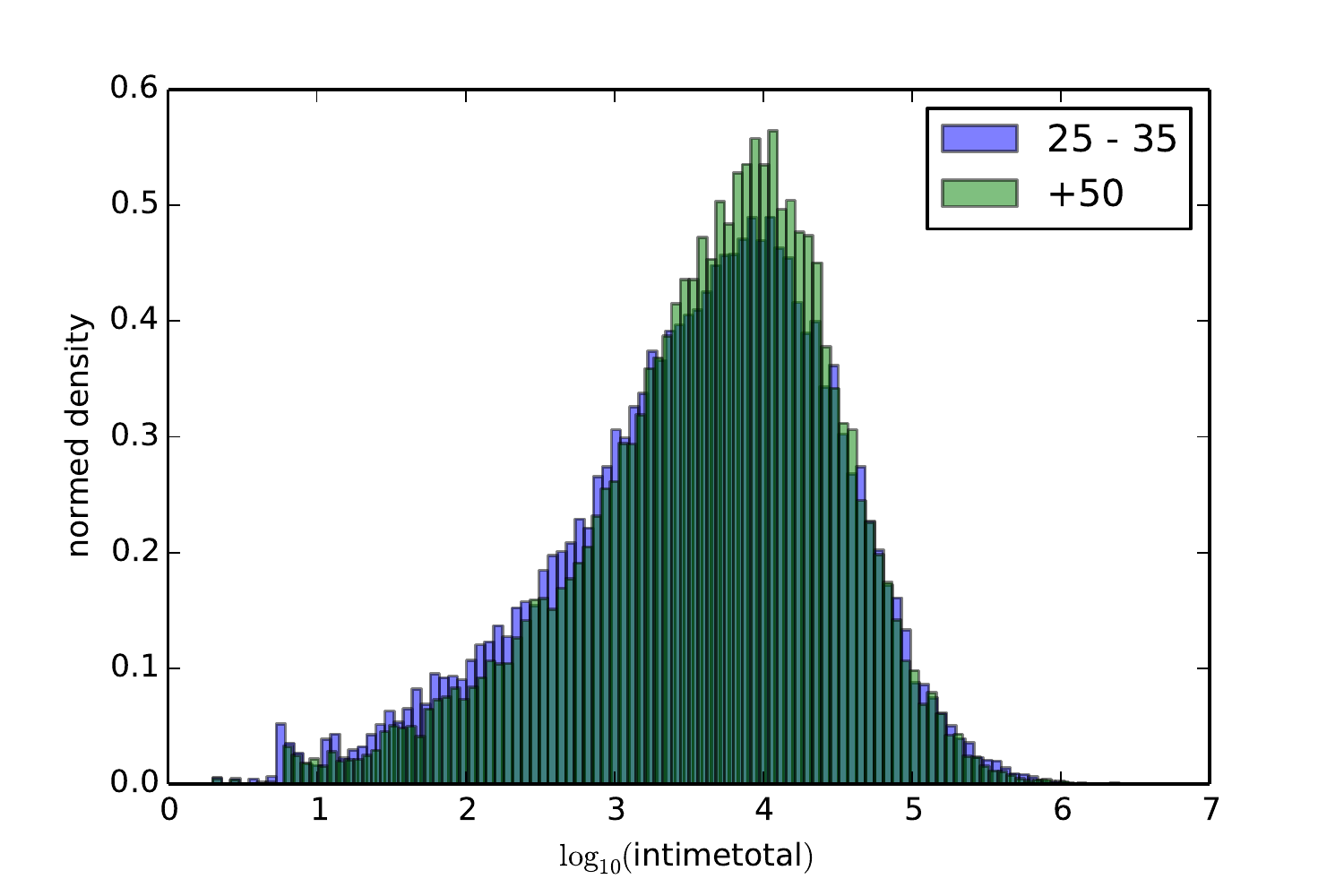}
		(e) 
	\end{minipage}
	\begin{minipage}{.48\linewidth}
		\centering
		\includegraphics[width=\textwidth]{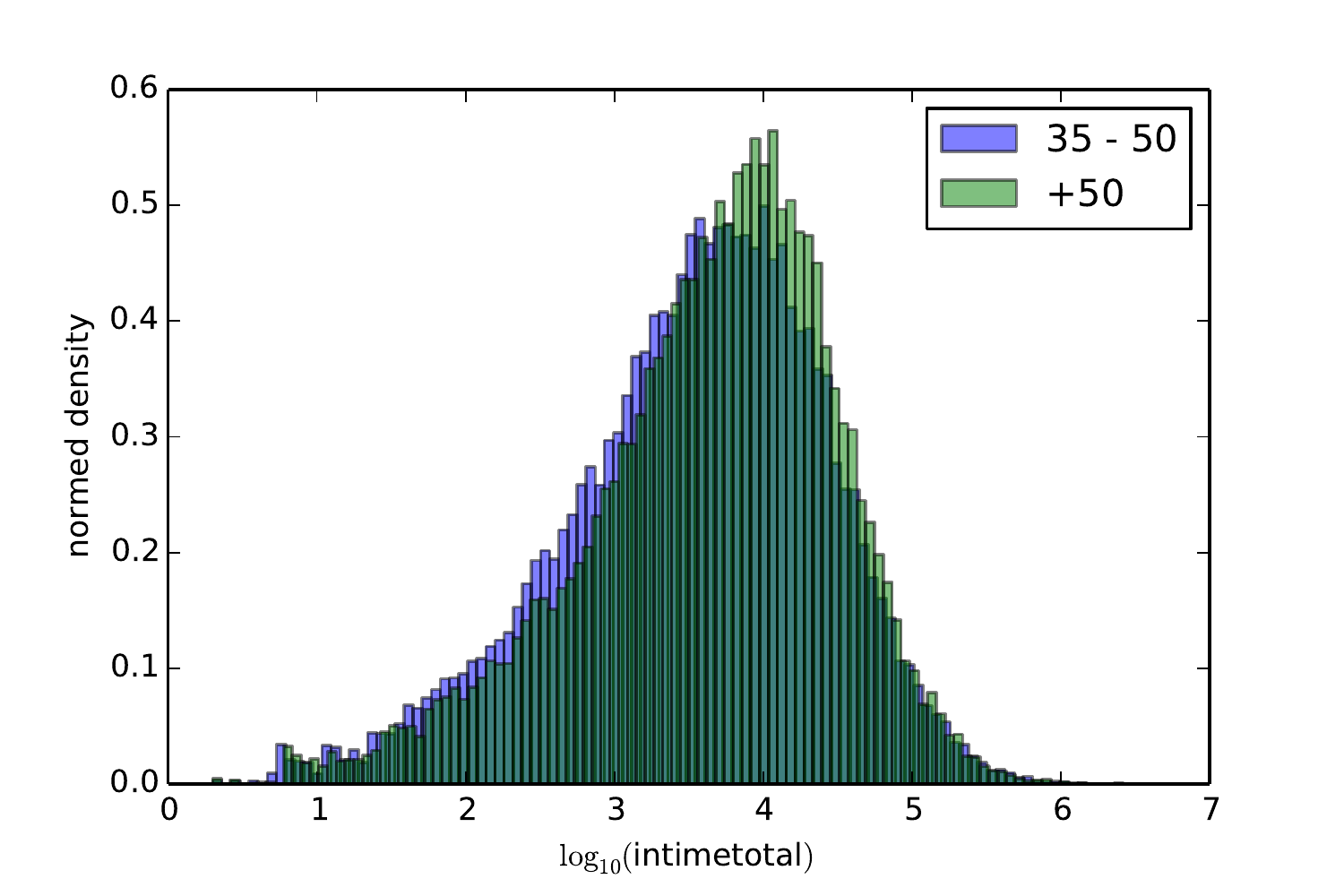}
		(f) 
	\end{minipage}
\end{footnotesize}

	\caption{Pairwise comparison of $\log_{10}(\textit{in-time-total} + 1)$  between different age groups.
	 The plots show the normed density for $\log_{10}(\textit{in-time-total} + 1) > 0$.}
	\label{fic:intimetotal_comparison}
\end{figure}

In Fig.~\ref{fic:intimetotal_comparison} we plot the distribution 
of $\log_{10}(\textit{in-time-total} + 1)$
for different pairs of age groups.
The following results can be observed from the plots:
\begin{itemize}
	\item The distribution for the group of people aged \textit{over 50} is shifted to the right in comparison with all the other age groups. This implies that people from this age group do talk more when called than people from any other age group. 
Figures~3c, 3e, 3f.
	\item The distribution for the group of people aged \textit{below 25} is shifted to the left. This distribution shows less kurtosis and a higher variance, meaning 
that this population is more spread in different levels of $\log_{10}(\textit{in-time-total} + 1)$. 
Figures~3a, 3b, 3c.
\end{itemize}

\subsection{Links between Age Groups} \label{sec:links-age-groups}

\begin{figure}[ht]
	\centering
	\includegraphics[width=0.95\linewidth]{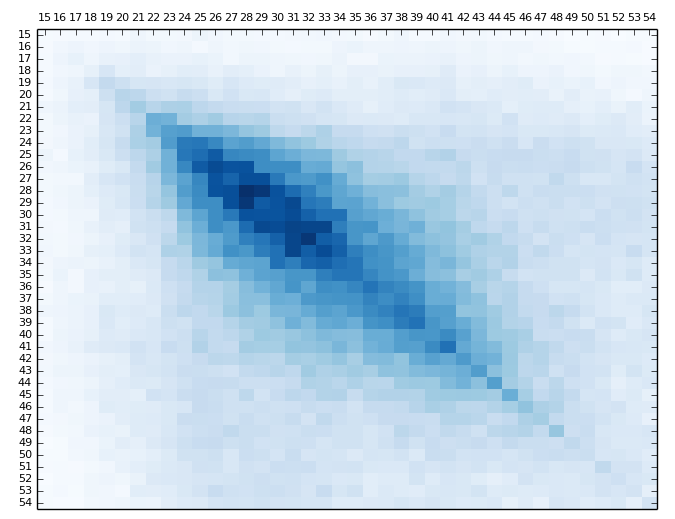}
	\caption{The matrix $C_{i,j}$ of communications between users of age $i$ and age $j$
	(the ages are expressed in years).}
	\label{fig:age-age}
\end{figure}

We study here the links between users, according to their age.
Fig.~\ref{fig:age-age} shows the matrix $C_{i,j}$ that contains the number of links between users of age $i$ and age $j$. 
For each user $u \in \calN_{GT}$, we compute the number of direct contacts of $u$ 
that belong to $\calN_{GT}$ and have age $j$. 
We sum over all the users of age $i$ to get the number $C_{i,j}$.
As we can see in the figure, the diagonal of the matrix has clearly higher values
than the rest of it, meaning that users are more likely to establish communications
with someone of their own age.
This strong \emph{age homophily} has also been observed in \cite{ugander5anatomy},
and in smaller social networks~\cite{mcpherson2001birds}.

\begin{figure}[ht]
	\centering
	\includegraphics[trim=0cm 0.2cm 0cm 0cm, width=0.95\linewidth]{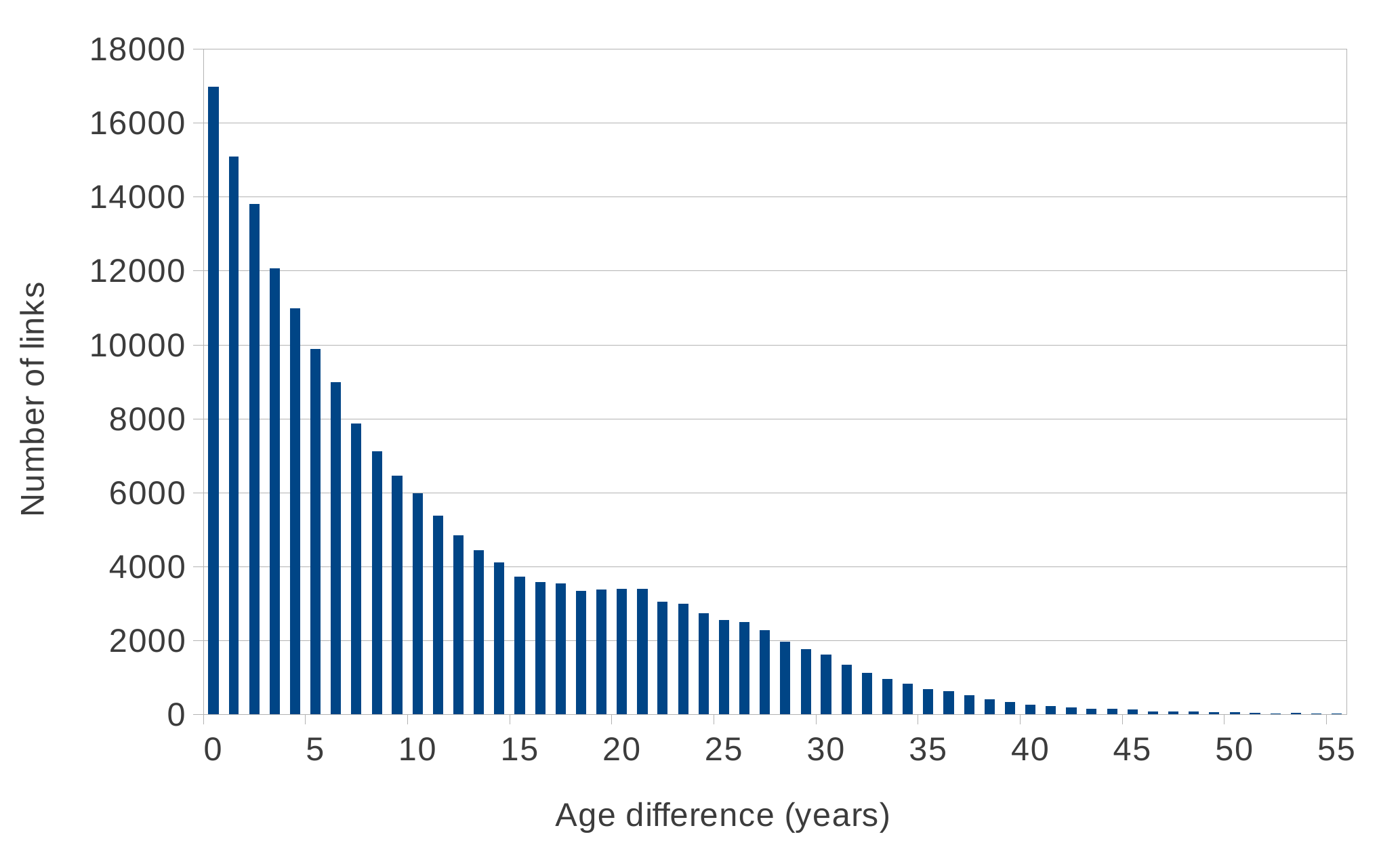}
	\caption{Number of links as a function of the age difference between users.}
	\label{fig:age-differences}
\end{figure}

The communication preferences can also be seen in Fig.~\ref{fig:age-differences},
which shows the number of links according to the age difference between users.
The highest number of links is observed when the difference is $\delta = 0$.
The number of links decreases with the age difference, except around the 
value $\delta = 21$, where an interesting inflection point can be observed;
possibly relating to different generations (e.g. parents and children).

\section{Age and Gender Prediction} \label{sec:identification}

This section describes the models that we used to estimate the age and gender of
users found in the dataset $\calN_O \setminus \calN_{GT}$. 
We show the results obtained using standard
Machine Learning models based on node attributes, 
applied to the prediction of gender (section~\ref{sec:gender-identification})
and age (section~\ref{sec:age-identification-node}).
We introduce a novel algorithm that leverages the communication network topology 
to generate age predictions (section~\ref{sec:age-identification-links}),
and finally show how it can be combined with the Machine Learning models (section~\ref{sec:age-identification-combined}).

We note that the feature variables are known for the whole set $\calN_O$,
while the target variables age and gender are known only for users 
in the set $\calN_{GT}$. We therefore use nodes belonging to $\calN_{GT}$
for our training and validation set, to predict both age and gender
for the remaining users in $\calN_O$.

\subsection{Population Pyramid Scaling (PPS)} \label{sec:pop-pyramid-scaling}

In the following subsections, we will present algorithms for gender and age prediction
which generate for each node, a probability vector over the possible categories
(gender or age groups).
At the end of their execution, when we ``observe'' the system, we are required to collapse
the probability vectors to a specific gender or age group. 
Choosing how to perform this collapse is not an obvious matter.
In effect, we want to collapse the probability state of the system as a whole and not for each node independently. 
In particular, we want to impose external constraints on our solution, 
namely that the gender or age group distribution for the whole network be that of the ground truth.
To achieve this, we developed a method that we call \emph{Population Pyramid Scaling}.
This algorithm takes, as a hyper-parameter, the proportion $q$ of
nodes to be predicted. For example, we use $q = 1/2$ to generate predictions
for the $50\%$ of nodes which got better classification results from the unconstrained method. 

\begin{algorithm}[t]
\ForEach{\emph{node $i$ and group $k$}}{
  compute the probability $p_{i,k}$ that node $i$ belongs to group $k$ 
  using the unconstrained algorithm\;
}
Create an ordered list $T$ of tuples $(i, k, p_{i,k})$\; 

Sort list $T$ in descending order by the column $p_{i,k}$.
The list $T$ will be iterated starting with the element with the highest probability\;

\ForEach{\emph{element $(i, k, p_{i,k}) \in T$}}{
		\If{\emph{node $i$ has not been assigned to a group}}{
		  \If{\emph{less than $N_k$ nodes assigned to group $k$}}{
		      assign node $i$ to group $k$\;
		  }
		}
}
\caption{Population Pyramid Scaling}
\label{algo:pps}
\end{algorithm}

The PPS procedure is described in Algorithm~\ref{algo:pps}.
Note that the population to predict has size $N = q \times | \calN_O |$.
For each category $k$ we compute the number of nodes $N_k$ 
that should be allocated to category $k$ in order to 
satisfy the distribution constraint (the gender or age distribution of $\calN_{GT}$),
and such that $\sum_{k = 1}^{C} N_k = N$ 
(where $C$ is the number of categories or groups).

\subsection{Gender Prediction} \label{sec:gender-identification}

For gender prediction, several algorithms were evaluated, with a preference for algorithms
more restrictive respect to the functions that they adjust.
Some of the algorithms used are:
Naive Bayes, Logistic Regression, Linear SVM, 
Linear Discriminant Analysis and Quadratic Discriminant Analysis.
As previously described, as part of data preprocessing, log transformation of the variables are added 
and the values are standardized to the $[0,1]$ interval.

\begin{table}[ht]
\caption{Best classifiers configuration for Gender Prediction}
\label{tab:classifier}
\renewcommand{\arraystretch}{1.2}
\centering
\begin{small}
\begin{tabular}{| l | l |}
\hline
Algorithm	& Best configuration \\
\hline
LinearSVC	& dual = True; penalty = L2;	 \\
			& loss = L1; C = 1; k = 100;		 \\
			& $|$ training set $|$ = 200,000 \\
\hline 
LogisticRegression 	& penalty = L1; C = 10; k = 100;				\\
			& $|$ training set $|$ = 200,000 \\
\hline
\end{tabular}
\end{small}
\end{table}

The best results were obtained with Linear SVM and Logistic Regression.
Table~\ref{tab:classifier} summarizes the classifiers configuration.
To find these parameters, we used grid search over a predefined set of parameters. 
For instance, the parameter $C$ for Logistic Regression takes its values
in the set $\{ 0.1, 0.3, 1, 3, 10 \}$.
Different number of attributes were evaluated before
training the model ($k \in \{10, 30, 100\}$).
The labeled nodes were split in a training set ($70\%$) 
and a validation set ($30\%$).

\begin{table}[ht]
\caption{Precision obtained for Gender Prediction}
\label{tab:results-gender-prediction}
\centering
\begin{small}
\begin{tabular}{  l  l l l l }
\toprule
Parameter $q$ & 1 & 1/2 & 1/4 & 1/8 \\
\midrule
Accuracy &  66.3\% & 72.9\% & 77.1\% & 81.4\% \\
\bottomrule
\end{tabular}
\end{small}
\end{table}

After performing PPS (to ensure the correct proportion of men and women), we obtained
the results shown in Table~\ref{tab:results-gender-prediction}.
As expected, the accuracy of our predictions improve when we decrease the parameter $q$,
which provides a trade-off between precision and coverage.
We reach a precision of 81.4\% when tagging 12.5\% of the users.

In the following paragraphs, we briefly recall some details of the classifiers that
gave the best results, in order to clarify the meaning of the configuration
parameters in Table~\ref{tab:classifier}. Those parameters are the ones required 
by the {Scikit-learn} library~\cite{scikit-learn:_2011}. 
In addition, 
{Pandas} \cite{mckinney-proc-scipy-2010} 
and {Statsmodels} \cite{statsmodels2010} 
have been used for the exploratory analysis.

\subsubsection{Linear SVC}\label{sec:linear_svc}

Classification using Support Vector Machines (SVC) requires optimizing the following 
function \cite{hsieh_dual_2008}:
\begin{equation}\label{eq:svc}
	\min_\omega \frac{1}{2}\omega^T\omega + C\sum_{i=1}^l \xi (\omega, x_i, y_i)
\end{equation}
where $y_i$ is the gender value and $x_i$ is the vector of normalized observed variables;
$\omega$ describes the hypothesis function,
and $C$ is the regularization parameter. 
In our case, we choose to use
$ \xi (\omega, x_i, y_i) = \max(1-y_i, \omega^Tx_i, 0) $.
This setup is called L1-SVM as $\xi(\cdot)$ defines the loss function.  
Implementation details can be found in 
\cite{scikit-learn:_2011,hsieh_dual_2008,fan_liblinear:_2008}.

\subsubsection{Logistic Regression}

A standard way to estimate discrete choice models is using index function models. 
We can specify the model as 
$ y^* =\omega x + \epsilon $,
where $y^*$ is an unobserved variable. 
We use the following criteria to make a choice:
\begin{align*}
\begin{cases}
	y = 1 = \text{Female}\quad & \text{if } y^* > 0\\
	y = 0 = \text{Male}\quad & \text{if } y^* < 0
\end{cases}
\end{align*}
Additionally, we use L1 regularization (for feature selection and to reduce overfitting). 
The complete  formulation is to optimize:
\[
	\min_\omega \|\omega\|_1 + C\sum_{i=1}^l \xi (\omega, x_i, y_i)
\]
where
$ \xi (\omega, x_i, y_i) = \log(1+\exp(-y_i\omega^Tx_i)) $.
For more information refer to
\cite{scikit-learn:_2011,fan_liblinear:_2008,greene_econometric_2011}.

\subsection{Age Prediction using Node Attributes} \label{sec:age-identification-node}

We now tackle the problem of detecting the age of the users using the properties of the nodes (users). 
As a first approach, we use the Machine Learning armory to perform the detection.
In order to reduce the complexity of the target variable, 
we partition it into $C$ age categories ($C = 4$): below 25 years old, from 25 to 34 years old, from 35 to 49 years old, and 50 years old or above (as in section~\ref{sec:observations-age}).
Given this set of categories, we found the best results using Multinomial Logistic (MNLogistic). This method is a generalization of Logistic Regression for the case of multiples categories 
(refer to \cite{statsmodels2010,greene_econometric_2011}).

A problem we encountered when using MNLogistic is the overfitting to the categories with higher frequencies, in this case the classification in categories 25 to 34 years and 35 to 49 years
of more elements than expected. 
In effect, in our Training Set, the age groups have the following distribution:
\smallskip
\begin{center}
\begin{small}
\begin{tabular}{  l  l l l l }
\toprule
Age group & 10-25 & 25-35 & 35-50 & 50+ \\
\midrule
Population &  12.1\% & 35.45\% &  37.45\% & 15\% \\
\bottomrule
\end{tabular}
\end{small}
\end{center}
\smallskip
But when using the MNLogistic algorithm, 
we obtained predictions with the following distribution:
\smallskip
\begin{center}
\begin{small}
\begin{tabular}{  l  l l l l }
\toprule
Age group &  10-25 & 25-35 & 35-50 & 50+\\
\midrule
Population & 0.66\% &  52.97\% &  45.52\% &  0.84\% \\
\bottomrule
\end{tabular}
\end{small}
\end{center}
\smallskip
To solve this issue we used the PPS (Population Pyramid Scaling) method of 
section~\ref{sec:pop-pyramid-scaling}.
After performing PPS, we obtained the results presented in Table~\ref{tab:results} (in 
section~\ref{sec:results-summary}).

\subsection{Age Prediction using Network Topology} \label{sec:age-identification-links}

As discussed in section~\ref{sec:links-age-groups}, there is a strong age homophily 
among nodes in the communication network, 
yet the Machine Learning algorithms we have employed so far
are mostly blind to this information, that is, their predictive power relies
solely on user attributes ignoring the complex interactions given by the mobile
network they participate in.

In this section we propose an algorithm which can harness the information
given by the structure of the mobile network and in this way, leverage hidden
information as the homophily patterns (shown in Fig.~\ref{fig:age-age}).

\subsubsection{Communication Network Structure}

We briefly describe how we construct the communication network $\calG$.
To each user (phone number) we assign a node in the network, and to each pair
of users $x$ and $y$ communicating via voice calls or SMS we assign a link (noted $x \sim y$). 
We can also assign a weight $w_{x,y}$ to each link between user $x$ and user $y$,
expressing the number of communications, the total activity or any other
property of the link. 
In the present study we choose to set $w_{x,y} = 1$ if there is
any kind of communication between nodes $x$ and $y$, and $w_{x,y} = 0$ otherwise. 
We do this both to reduce the complexity of the analysis, and because as a first
approach, we are interested in exploring how the bare network topology enhances
our predictions of the target variable.

\subsubsection{Reaction-Diffusion Algorithm}
\label{sec:graph-based}

In our dataset, we have the values for the target variable (age)
of the nodes in $\calN_{GT}$, and
we can also see from Fig.~\ref{fig:age-age} that neighbouring nodes are more likely  
to belong to the same age category. 
A mathematical model that can take advantage of this information
together with the topology of the network to infer the target
values for the remaining nodes (in $\calN_O \setminus \calN_{GT}$) is that of a 
diffusion process in a graph. 
At each time step the information (value of target variable at a given node) is diffused or
spread to its neighbours. In this way, given enough time steps, the information from
nodes in $\calN_{GT}$ is diffused to the entire network.
Now, if $| \calN_{GT} | \ll | \calN_O |$, which is the case in our study, pure diffusion may not be strong
enough to have the information in $\calN_{GT}$ significantly affect the values of the target
variables in the entire network. To remedy this, we include a reactive
term in our algorithm where at each time step, nodes in $\calN_{GT}$ are reinforced with
their value at time $t = 0$. Given that we have partitioned our target variable
in $C$ categories, we also found it advantageous to model our
reaction-diffusion process as one where the information being diffused is the
probability distribution for each node (noted $g_{x,t}$) to belong to each category. 
We detail the algorithm below.

For each user $x$ we define the initial state $f_{x} \in \mathbb{R}^{C}$ (where $C$ is the number of categories) as having components
\begin{equation}
	\left(f_{x}\right)_i = \begin{cases}
		\delta_{i,a(x)} & \text{if }  x \in \calN_{GT}  \\
		1/C   & \text{if } x \not\in \calN_{GT}
	\end{cases}
\end{equation}
where $a(x)$ is the age category of user $x$
and $\delta_{i,a(x)}$ is the Kronecker delta function.
Then we define $g_{x,t}$ as
\begin{equation}
\label{eq:difusiongeneral}
	\begin{split}
		g_{x,0} &= f_{x} \\ 	
		g_{x,t} &= (1-\lambda) \; f_{x}+ \lambda \; \frac{\sum_{x\sim y} w_{y,x} \, g_{y,t-1} }{\sum_{x\sim y}w_{y,x}}
	\end{split}
\end{equation} 
where $x \sim y$ means that there is a link between $x$ and $y$; $w_{y,x}$ is the weight of the link;
and $\lambda$ is a hyper-parameter which tunes the relative strength 
of the reinforcement and diffusion terms (which we set to $\lambda = 0.5$ in our experiments).
Note that $g_{x,t} \in \mathbb{R}^{C}$ is the discrete probability measure for node $x$ at time $t$,
in particular the sum 
$\sum_{i=1}^{C} (g_{x,t})_i = 1 \; \forall \, x, t$. 

As previously stated, these equations are similar to those of a \emph{reaction-diffusion} 
process (we note that it can also be seen as a Jacobi method for the appropriate linear system). 
For the experimental results, we consider a simpler model by taking $w_{y,x}=1 \; \; \forall x \sim y$
and $0$ otherwise. 
The equation for $g_{x,t}$ becomes
\begin{equation}
\label{eq:difusion1}
		g_{x,t} = (1-\lambda) \; f_{x}+ \lambda \; \frac{\sum_{x\sim y} g_{y,t-1} }{|\{y:x\sim y\}|} .
\end{equation}
We iterate this process $m$ times (for $1 \leq t \leq m$). In our experiments, $m = 30$ was sufficient for the process to converge.

For each $x$, we obtain a vector $g_{x} = g_{x,m}$.
With this we can get a prediction for the age of $x$ given by $argmax _{1\leq i \leq 4} (g_{x})_i$. 
This prediction, in contrast with the prediction performed the MNLogistic model 
(based on node attributes), gave us a population pyramid closer to the ground truth:
\smallskip
\begin{center}
\begin{small}
\begin{tabular}{  l  l l l l }
\toprule
Age group &  10-25 & 25-35 & 35-50 & 50+\\
\midrule
Population & 7.26\% &  32.42\% &  50.49\% &  10.36\% \\
\bottomrule
\end{tabular}
\end{small}
\end{center}
\smallskip
After adjusting the distribution with the PPS algorithm
(from section \ref{sec:pop-pyramid-scaling}),
we obtained the results shown in Table~\ref{tab:results}.

\subsection{Enriching the Graph Algorithm with Node Attributes} 
\label{sec:age-identification-combined}
We propose here an algorithm to predict the age of users that leverages 
the \emph{PPS} algorithm, the node classification of section~\ref{sec:age-identification-node}
and the pure graph-based \emph{Reaction-Diffusion} algorithm. 
We define as initial state:
\begin{equation}
	f_{x} = \begin{cases}
		\delta_{i,a(x)} & \text{if }  x \in \calN_{GT}  \\
		\text{ML}(x)   & \text{if } x \not\in \calN_{GT}
	\end{cases}
\end{equation}
where $\text{ML}(x)$ is the result given by the best \textit{Machine Learning} algorithm of section~\ref{sec:age-identification-node} 
(i.e. Multinomial Logistic). 

Then, as before, the iterative process follows Equation~\eqref{eq:difusion1}.
In this case, the hyper-parameter $\lambda$ provides a trade-off between the information from the network topology and the initial information obtained with Machine Learning methods over node attributes
(here again we take $\lambda = 0.5$).

\subsection{Summary of Results} 
\label{sec:results-summary}

Finally, Table~\ref{tab:results} summarizes the results obtained with the different methods:
Machine Learning (ML) alone, Reaction-Diffusion (RDif) alone, and the combined method (ML + RDif).
We report for each case the accuracy obtained, that is the percentage of correct 
predictions on the validation set.

\begin{table} [ht]
\caption{Precision obtained for Age Prediction}
\label{tab:results}

\centering
\begin{small}
\begin{tabular}{ c c c c }
\toprule
\textbf{Population} & \textbf{ML} & \textbf{RDif} & \textbf{ML + RDif}\\
\midrule
$q$=1~~ & 36.9\% & 43.4\% & 38.1\%	\\
$q$=1/2 & 42.9\% & 47.2\% & 46.3\%	\\
$q$=1/4 & 48.4\% & 56.1\% & 52.3\%	\\
$q$=1/8 & 52.7\% & 62.3\% & 57.2\%\\
\bottomrule
\end{tabular}
\end{small}
\end{table}

The table shows that taking a smaller $q$ improves the accuracy of the results.
Our experiments also show that the RDif (Reaction-Diffusion) algorithm outperforms 
the ML predictions based on node attributes.
It is also interesting to remark that the RDif algorithm outperforms
the combined method. 
The best precision obtained is 62.3\% of correctly predicted nodes, 
when tagging 12.5\% of the population.
Note that random guessing the age group (between 4 categories) would yield a precision of 25\%.

\section{Conclusion and Future Work} \label{sec:conclusion}

To our knowledge, this work provides the first extensive study of
social interactions in the country of Mexico focusing on gender and
age, based on mobile phone usage.
From a sociological perspective, the ability to analyze
the communications between tens of millions of people allows us to make strong inferences
and detect subtle properties of the social network.

As described in section~\ref{sec:observations}, the graph we constructed has
very rich link semantics, containing a detailed description of the communication patterns (45 characterization variables).
With PCA, we found that most of the variance of the characterization variables is contained in
a low dimensional subspace.
Motivated by these results, we focused on how the statistical properties of
the most informative attributes vary with both gender and age. 
In section~\ref{sec:observations-gender}, 
we make two interesting observations: 
(i) there is a gender homophily in the communication network (see Equation~\ref{eq:gender-homophily});
(ii) an asymmetry respect to incoming and outgoing calls can be observed between men and women,
possibly reflecting a difference of roles in Mexican society (it would be interesting to see
how these differences change in other regions like Europe or the United States).
We also compared communication habits for different age groups,
and found statistically significant differences.
Finally, our most important observational contribution is the study of correlations
between age groups in the communication network, as summarized in Fig.~\ref{fig:age-age}
and \ref{fig:age-differences}.
We observe a strong age homophily \cite{mcpherson2001birds},
and a strong concentration of communications centered 
around the age interval between 25 and 45 years. 
But we also notice weaker modes in both figures,
which raise interesting sociological questions
(e.g. whether they reflect a generational gap).

The second key contribution of this work was to study and propose novel methods
to infer the gender and age of users in the mobile network.
As a first approach, described in sections~\ref{sec:gender-identification} and \ref{sec:age-identification-node}, we used a set of standard Machine Learning tools finding that Logistic Regression 
and Linear Support Vector Machine algorithms gave us the best results.
However, these techniques cannot harness the topological information
of the network to explore possible correlations between the users' age groups. 
To leverage this information, we proposed an purely graph based algorithm inspired in a 
\emph{reaction-diffusion} process, and demonstrated that with this methodology we could 
predict the age category for a significant set of nodes in the network. 
Our experiments showed that the \emph{reaction-diffusion} method provides the best predictive power
on a real-world large scale dataset.

There are multiple directions in which this work can be extended.
We highlight the following:

\paragraph{Analysis of Hyper Parameters}
The analysis of the prediction performance as a function of the hyper-parameters $q$ and $\lambda$, 
used in sections \ref{sec:age-identification-node} 
and \ref{sec:age-identification-links}, is important for a fine tuning of the algorithm. 
We are also interested in studying the effect of variations in the weights $w_{x,y}$ used in 
the diffusion process
(e.g. use the intensity of communication or the geolocation data to weight the links).
In particular, we want to explore how the network topology information can be combined with 
nodes features to improve the joined (ML + RDif) methodology.

\paragraph{Extend Depth}
A statistics quasi-experiment can be built from this method \cite{william2002experimental}.
In this case, we want to know whether the differences in the observed behavior can be accounted 
to gender and age, or are consequences of differences in the ego-network induced by phone calls. 
This quasi-experiment can be performed using Propensity Score \cite{rosenbaum1983central},
and may provide sociological insights.

\paragraph{Extend Width}
One direction that we are currently investigating is to apply the methodology presented here
to predict variables related to the users' spending behavior.
In \cite{singh2013predicting} the authors show correlations between social features
and spending characterizations, for a small population (52 individuals).
We are interested in applying our methodology to predict spending behavior characteristics
on a much larger scale (millions of users).

Another research direction is to use the geolocation information contained in the Call Details Records.
Recent studies have focused on the mobility patterns related to cultural events
--for instance sport related events \cite{ponieman2013human,xavier2013understanding}--
which might exhibit differences between genders and age groups.
Looking at mobility patterns through the lens of gender and age characterization will provide
new features to feed the Machine Learning part of our methodology,
and more generally will provide new insights on the human dynamics of different segments of the population.


\bibliographystyle{IEEEtran}

\bibliography{sna}

\end{document}